\documentclass[largeformat]{interact}
\pdfoutput=1

\usepackage{subfigure}% Support for small, `sub' figures and tables
\usepackage[pdftex,colorlinks=true,bookmarks=false,citecolor=blue,urlcolor=blue]{hyperref}

\usepackage[numbers,sort&compress,merge]{natbib}% Citation support using natbib.sty
\bibpunct[, ]{(}{)}{,}{n}{,}{,}% Citation support using natbib.sty
\theoremstyle{plain}% Theorem-like structures

\theoremstyle{definition}

\theoremstyle{remark}

\begin{document}

%\articletype{ARTICLE TEMPLATE}

\title{Controlled Supercontinua via Spatial Beam Shaping}

\author{
\name{Alexandra A. Zhdanova\thanks{CONTACT A.~A. Zhdanova. Email: sashaa@physics.tamu.edu}, Yujie Shen, Jonathan V. Thompson, Marlan O. Scully, Vladislav V. Yakovlev, Alexei V. Sokolov}
\affil{Institute for Quantum Science and Engineering, Department of Physics and Astronomy, Texas A \& M University, College Station, TX 77843-4242 USA}
}

\maketitle

\begin{abstract}
	
Recently, optimization techniques have had a significant impact in a variety of fields, leading to a higher signal-to-noise and more streamlined techniques. We consider the possibility for using programmable phase-only spatial optimization of the pump beam to influence the supercontinuum generation process. Preliminary results show that significant broadening and rough control of the supercontinuum spectrum are possible without loss of input energy. This serves as a proof-of-concept demonstration that spatial effects can controllably influence the supercontinuum spectrum, leading to possibilities for utilizing supercontinuum power more efficiently and achieving arbitrary spectral control.

\end{abstract}


\section{Introduction}

Stable broadband sources are essential for a variety of applications, from spectroscopy \cite{Liu2013} to imaging \cite{Kano2005} and nonlinear optical parametric amplifiers \cite{Reed1995}. Supercontinuum (also known as white light) generation offers a promising tool for these multidisciplinary uses. Sapphire is regarded as the crystal of choice for visible supercontinuum generation \cite{Yakovlev1994} and is the focus of our particular study. Supercontinua are generated in sapphire through the process of single filamentation. This general process encompasses a wide variety of linear and nonlinear effects including self-steepening, self-phase modulation, dispersion, four-wave mixing, Raman excitation, second and third harmonic generation, and plasma generation, absorption, and refraction \cite{Alfano2006}. Since in the most general case it is diffcult to separate the exact contribution of each of these, individual optimization of each effect is neither particularly feasible nor extremely desirable. However, each of these effects can be highly influenced by the spatial distribution of the beam. One of the most basic effects of spatial shaping is to focus the beam tighter, thereby generating more plasma and influencing the resultant spectrum. Although studies with Bessel \cite{Majus2013} and Laguerre-Gauss beams \cite{Sztul2006a} have failed to find significant spectral deviations or improvements in effciency from the Gaussian regime, systematic spatial optimization has not been performed. 

Previous work that has shown promise in this direction includes a study performed with microlenses generated via spatial light modulators \cite{Borrego-Varillas2014}. Moreover, there have been several successful studies of spectral pump-beam optimization in filamentation \cite{Thompson2017, Ackermann2006} and other nonlinear effects, such as the control of second harmonic generation \cite{Thompson2016} and the enhancement of spontaneous Raman signals through a turbid medium \cite{Thompson2016a}. We extend these results and methods to the theoretically challenging regime of supercontinuum generation by using the wavefront shaping algorithm to influence the supercontinuum spectrum.

There are several aspects of interest here, including reshaping the physics of filamentation and drawing attention to the potentials of non-Gaussian beams. More commercial applications are also possible. We envision that this work may lead to universally stronger seeds for spectroscopic applications that depend on nonlinear effects for a large signal-to-noise. Our preliminary results are a proof-of-concept that the idea of spatial optimization has merit and may be further expanded with additional work.

\section{Experimental Setup}

\begin{figure}%[htbp]
	\centering

	%width=1\textwidth,height=50mm
	\includegraphics[width=0.5\textwidth]{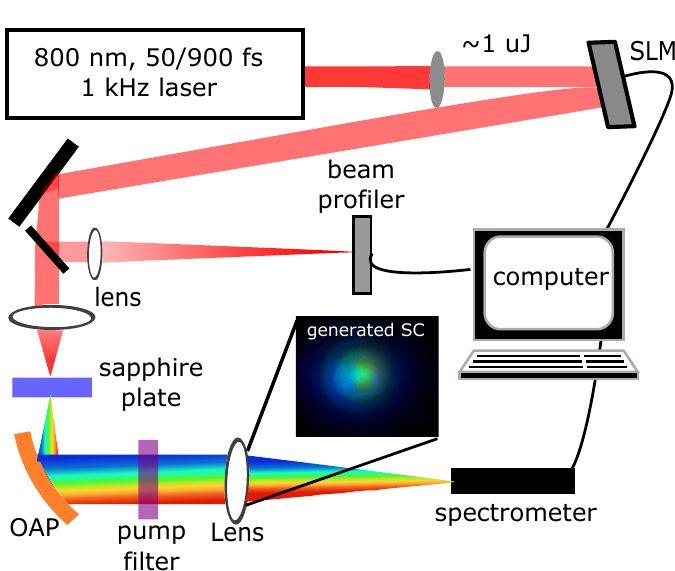}
	\caption{Setup for generating supercontinua (SC) from shaped pulses; a photograph of the generated SC is shown in the inset. The angle of the spatial light modulator (SLM) is greatly exaggerated. OAP stands for off-axis parabola and was used to collimate the SC spectrum after the crystal.}
	\label{fig:setup}

\end{figure}

The experimental setup is depicted in Fig. \ref{fig:setup}. We used a Ti:Sapphire regenerative amplifier (Coherent, Legend) to produce infrared ($\lambda = 802$ nm) $35$ fs pulses with a $1$ kHz repetition rate and $4$ W average power that we attenuated to produce supercontinuum. We then investigated two regimes of supercontinuum generation: chirped and unchirped. In the first case, we added positive chirp by changing the grating distance within the compressor unit of the amplifier. This produced pulses of 900 fs FWHM duration, measured using a commercial autocorrelator (Pulse Check; A.P.E.).

Different powers/pulse were needed to produce supercontinua in each regime: we used 5-6 $\mu$J in the chirped and 1 $\mu$J in the unchirped cases. In both regimes, we used a spatial phase-only light modulator (Hamamatsu; x10468; abbreviated SLM) to shape the originally Gaussian beam. The beam has a 15 mm diameter prior to being shaped by the SLM; this is the greatest we could expand the beam without clipping on the SLM screen. We then split the beam and focused part of it with a 5 cm focal length lens to generate a supercontinuum in a 3 mm thick single crystal c-cut sapphire plate (Newlight Photonics; SAP0030-C; Toronto, ON). The resultant diverging supercontinuum was collimated by an off-axis paraboloid to a 1-inch diameter. The pump beam was then filtered out via a 750 nm short pass filter (Semrock; FF01-745/SP-25; Rochester, NY). The filtered beam was subsequently refocused with a 20 cm focal length lens into a multi-mode 600 $\mu$m core fiber. Since the supercontinuum light should roughly focus to 5 $\mu$m, the fiber core is of sufficient size to collect all the light and not be affected by any spatial phase changes the SLM adds. We checked this assumption by translating the fiber in x-y dimensions in the focal plane; this operation revealed no unmeasured light. Hence, we are confident that the VCSA algorithm is not optimizing the light-collection system.

The other part of the beam was sent through a long (2 m focal length) lens to be very loosely focused onto a CCD array (Spiricon; SP620U) and recorded by the computer. These images did not take part in the spatial optimization at all -- they are there to help visualize the effect of different phase maps on the beam's spatial profile in the focus. 

\section{Optimization Details}

For all optimization regimes, we used a variation of the continuous sequential algorithm (abbreviated VCSA) \cite{Vellekoop2008, Thompson2015}. The VCSA groups pixels on the SLM together and cycles through $2\pi$ phase values in 8 steps. The algorithm then compares spectrometer output in a particular spectral range before and after adding different phase values. If the average of the spectrometer reading in that spectral range improves, then the algorithm keeps the phase value. This cycle is repeated three times and results averaged to minimize influence from shot-to-shot fluctuations and other noise. The algorithm then moves on to another pixel group and repeats the process. Each iteration takes 12 seconds, with the spectrometer integration time forming the largest limit on speed. 

\begin{figure}%[htbp]
	\centering

	%width=1\textwidth,height=50mm
	\includegraphics[width=0.5\textwidth]{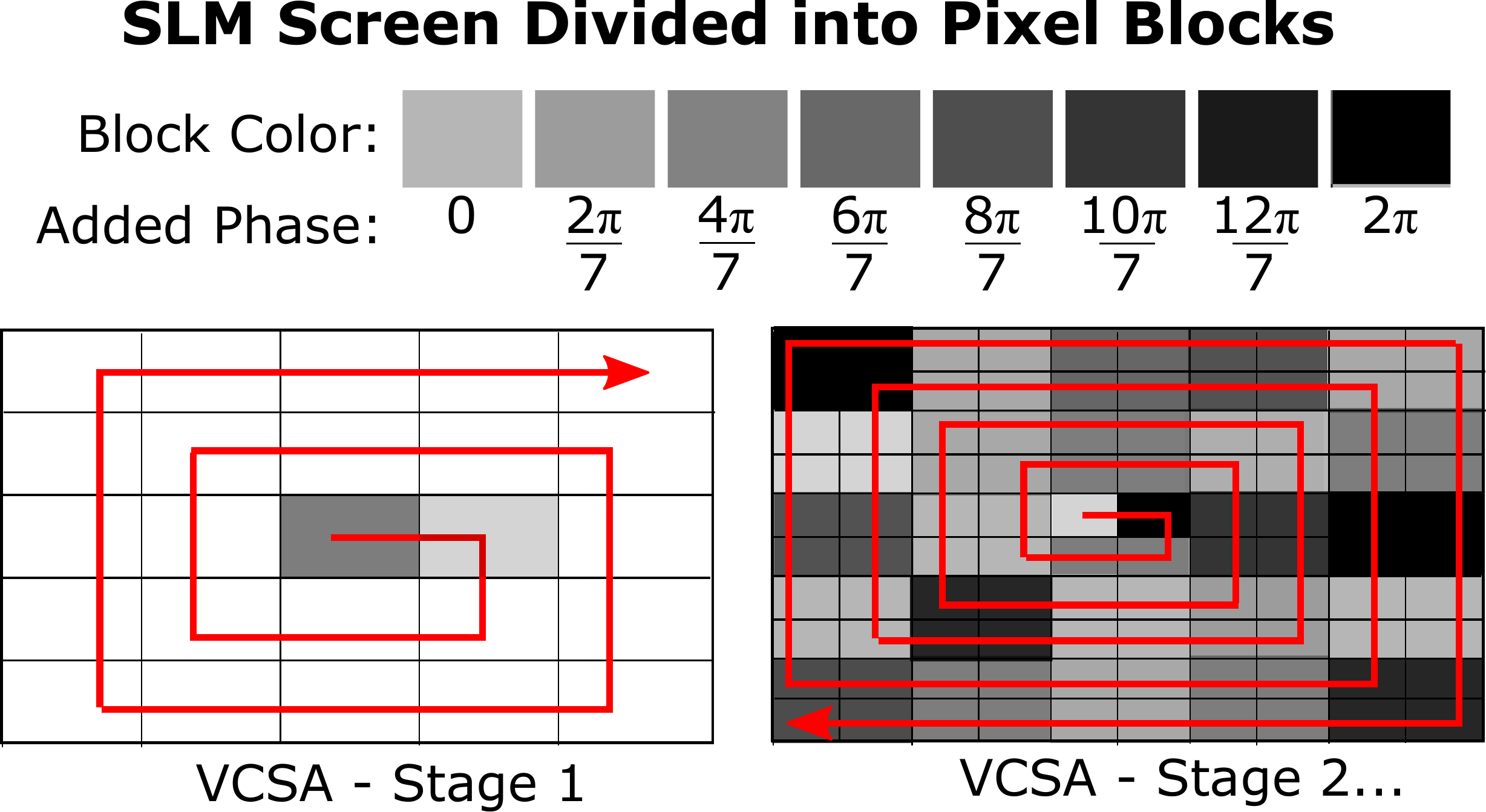}
	\caption{The smaller the block the longer the program takes to finish; the process can be stopped at any time if the user is satisfied with the results. Our SLM is comprised of 792 $x$ 600 pixels total.}
	\label{fig:algorithm}

\end{figure}

For all results given in this paper, we employed the ``spiral out" method of this algorithm, which starts with large pixel groups (of 264 $x$ 300 pixels) in the center and spirals out to the edges, as in Figure \ref{fig:algorithm}. It then starts a new stage at the center with smaller pixel groups (of 132 $x$ 150 pixels) and spirals out until it is forced to repeat itself with even smaller groups of pixels (of 72 $x$ 60). The final run consists of groups of 24 $x$ 24 pixels. In total, we let the algorithm optimize for roughly half an hour for the results given in this paper. We do not consider time to be a major limit in our experiment, as there are no discernible differences between spectra taken at the beginning of the day and those taken at the end. Further, for spectroscopic applications, it will not be necessary to quickly reoptimize the masks so long as the user takes care to produce a bank of working masks that they may easily switch between.

\section{Preliminary Results}

\begin{figure}%[htbp]
	\centering

	%width=1\textwidth,height=50mm
	\includegraphics[width=0.5\textwidth]{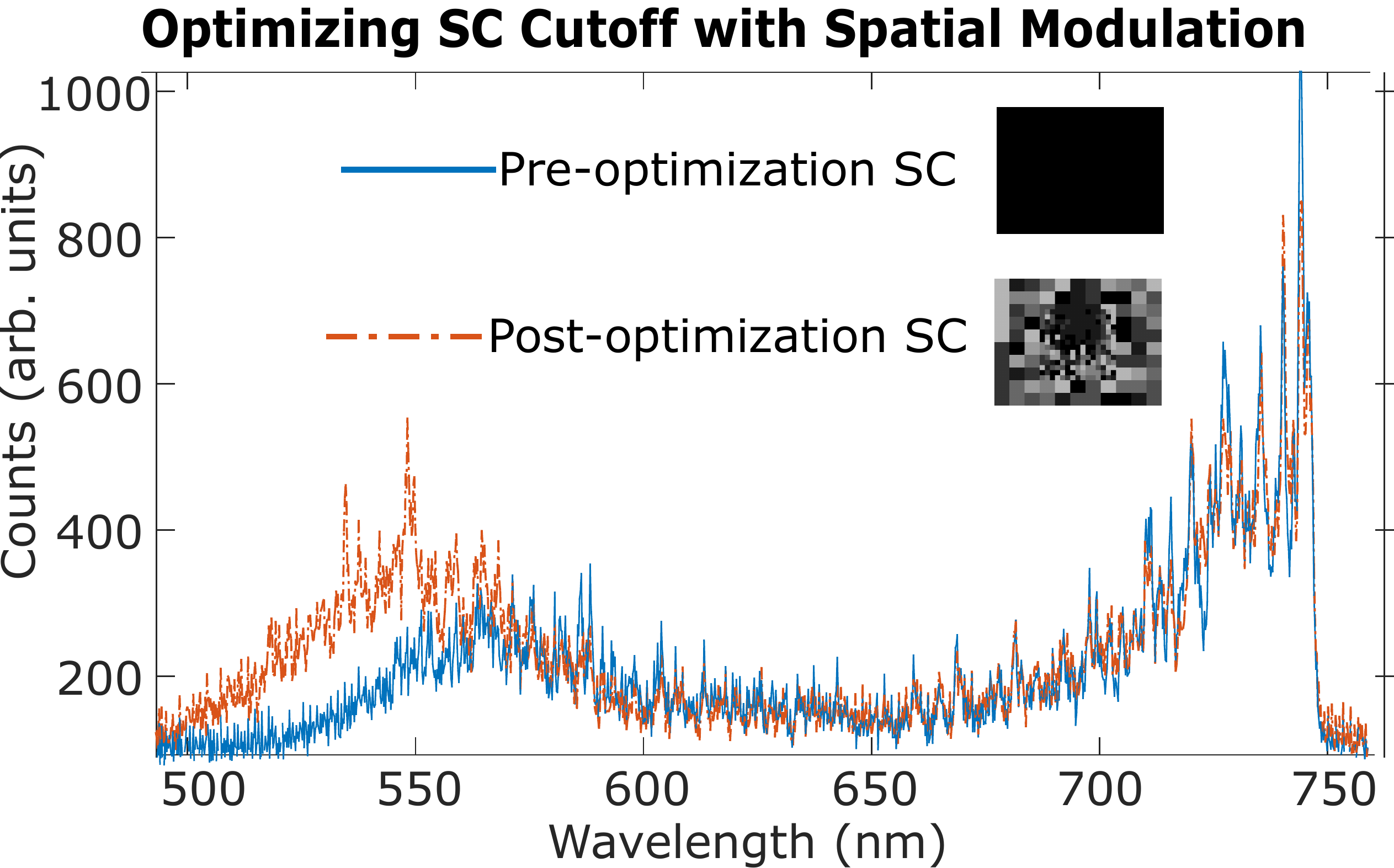}
	\caption{SC spectrum before (blue line) and after (red line, dash-dot) spatially optimizing the pump pulse. The range of optimization was 500--550 nm.}
	\label{fig:chirpedresults}

\end{figure}

Using these methods, we were able to obtain a general 10\% broadening of the spectral width of the supercontinuum generation for highly positively chirped pulses (900 fs), as shown in Fig. \ref{fig:chirpedresults}. However, this regime is tricky to work with as the damage threshold for these focusing conditions in sapphire is near the critical power of self-focusing. This makes further optimization difficult but not impossible, potentially under different focusing conditions.

\begin{figure}%[htbp]
	\centering

	%width=1\textwidth,height=50mm
	\includegraphics[width=1\textwidth]{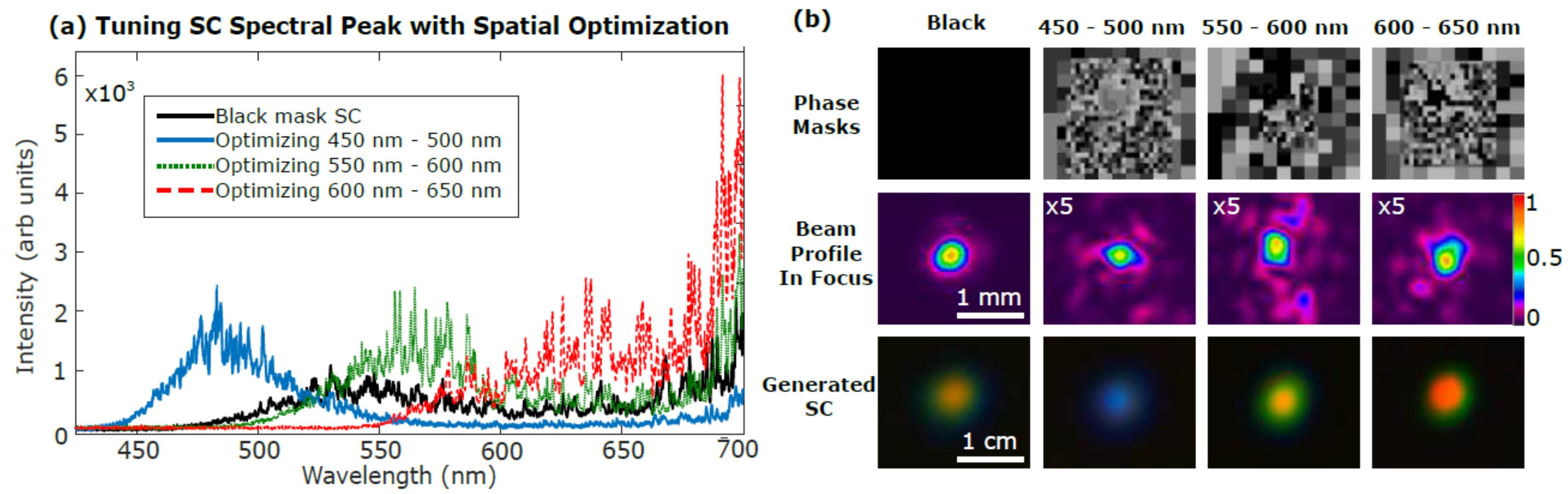}
	\caption{(a) Measured supercontinuum spectrum for different optimization regimes -- the supercontinuum cutoff peak is spectrally shifted as the spatial shape changes. Each entry in the legend corresponds to the optimization range of that particular run of the algorithm (i.e. for the second entry, the algorithm attempts to optimize the average spectrometer-measured counts in the range of 450 -- 500 nm). All spectrums were taken with the phase masks and profiles in (b). (b) SLM phase masks (top), beam profiles in the focus magnified approximately 40 times and with the left three profiles integrated 5x longer than the right-most profile (middle), and true-color photographs of the resultant supercontinuum (bottom; taken with a Sony DSLR camera) for different optimization regimes.}
	\label{fig:results}

\end{figure}

For 35 fs unchirped pulses, we discovered that it is possible to shift the supercontinuum spectral cutoff peak between 450 and 650 nm, as our preliminary results indicate in Figure  \ref{fig:results}. The region from 450-500 nm is completely absent in the supercontinuum spectrum generated without any phase mask applied and so represents a significant broadening ($>\approx 20\%$). In this case, the effect of the added phase mask on the supercontinuum spectrum is easily noticeable by eye and hence can not be due to any limitations in our light-collection system.

 Further, the phase masks shown in Figure \ref{fig:results} generate the same spectrum from day-to-day without any special additional environmental control, making our experiment repeatable in a variety of conditions. However, the spatial profile of the shaped beam is very sensitive to the alignment of the pump beam on the SLM screen. This is because any displacement in this region will result in different parts of the beam obtaining different phase values, and hence not reproducing the original phase-optimized beam. In this case, each phase mask will need to be re-optimized to obtain a tailored spectrum.

\section{Conclusions}

Our preliminary results indicate that spatial beam shaping has a substantial untapped potential in optimizing supercontinuum generation by enhancing a particular spectral region. We envision that this technique can dramatically improve the ability to tailor the supercontinuum spectrum for any particular application. For example, we can provide significantly stronger seed pulses for optical parametric amplifiers and substantially enhance signals in broadband coherent anti-Stokes Raman spectroscopy/microscopy. An SLM provides a much more flexible platform, as compared to a micro-structured fiber, to tailor the spectral properties of the supercontinuum
\cite{Zheltikov2006}. The user will simply load the SLM with the phase mask for the particular spectral range they desire. By pre-generating optimal phase masks, the frequency can be tuned at the 10 Hz refresh rate of the SLM.

In the future, we envision that this will lead to higher available powers for various nonlinear spectroscopy experiments and hence a greater signal-to-noise, paving the way for future precision measurements. Further work will include explorations of the theoretical foundations of spatial effects in high-order nonlinear optical interactions, which we initiated in \cite{Thompson2016}. We also plan thorough investigations of algorithmic shaping in the IR.

This research was partially supported by the NSF (Grant \# PHY-1307153, DBI-1532188, and ECCS-1509268), the US Department of Defense (award FA9550-15-1-0517), the Welch Foundation (Awards No. A-1547 and No. A-1261), the Cancer Prevention Research Institute of Texas (grant RP160834), and the Office of Navel Research (Award No. N00014-16-1-3054).

\bibliographystyle{jmo_test2}
\bibliography{sc_jmo}
\end{document}